\documentclass{aipproc}
\layoutstyle{6x9}
\usepackage{bm}
\begin{document}

\title{The heavy-quark hybrid meson spectrum in lattice QCD}


\author{K.~Jimmy Juge}{
  address={Institute for Theoretical Physics, University of Bern,
           Sidlerstrasse 5, CH-3012 Bern, Switzerland}}
\author{Julius Kuti}{
  address={Department of Physics, University of California at San Diego,
           La Jolla, USA  92093-0319}}
\author{Colin Morningstar}{
  address={Department of Physics, Carnegie Mellon University, Pittsburgh,
           PA, USA  15213-3890}}

\begin{abstract}
Recent findings on the spectrum of heavy-quark mesons from computer
simulations of quarks and gluons in lattice QCD are summarized,
with particular attention to quark-antiquark states bound by an
excited gluon field.  The validity of a Born-Oppenheimer treatment
for such systems is discussed.  Recent results on glueball masses,
the light-quark $1^{-+}$ hybrid meson mass, and the static three-quark
potential are summarized.
\end{abstract}

\maketitle

\section{Introduction}

Much of our current understanding of hadron formation comes from the
constituent quark model.  The quark model is motivated by quantum
chromodynamics (QCD) and views hadrons as valence quarks interacting
via an instantaneous confining Coulomb plus linear potential.  In
the quark model, the gluons are recognized as the source of the
confining potential, but their dynamics is completely ignored.

Most of the observed hadron spectrum is described reasonably well
by the quark model.  The agreement is remarkable given
the crudeness of the model.  
In the quark model, mesons may have only certain $J^{PC}$ quantum
numbers: if the total spin and orbital angular momentum of the
quark-antiquark pair are $S=0,1$ and $L=0,1,2,\dots$, respectively, then
the parity and charge conjugation are given by 
$P=(-1)^{L+1}$ and $C=(-1)^{L+S}$.  In other words, 
$0^{+-},0^{--},1^{-+},2^{+-},3^{-+},4^{+-},\dots$ are forbidden
and mesons having such $J^{PC}$ are known as exotics.  Both an
overabundance of observed states and recent observations of 
exotic $1^{-+}$ resonances\cite{E852}
underscore the need to understand hadron formation beyond the
quark model.

QCD suggests the existence of states in which the gluon field is
excited.  Such states with no valence quark content are termed
glueballs, whereas states consisting of a valence quark-antiquark
pair or three valence quarks bound by an excited gluon field are known
as hybrid mesons and hybrid baryons, respectively.  Glueballs and
hybrids are
currently not well understood, making their experimental identification
difficult.  Theoretical investigations into their nature must confront
the long-standing problem of dealing with nonperturbative gluon field
behavior, but for this reason, such states are a potentially rich
source of information concerning the confining properties of QCD.

In this talk\footnote{Presented by C.~Morningstar.}, progress in
understanding heavy-quark conventional and
hybrid mesons using lattice simulations of gluons is reported.  
The validity of a Born-Oppenheimer treatment of such
systems is discussed.  Results on glueball masses, the light quark
$1^{-+}$ hybrid meson mass, and the static three-quark potential are
also summarized.

\section{Heavy-quark hybrid mesons}

The study of heavy-quark mesons is a natural starting point in the
search to understand hadron formation.  The vastly different characteristics
of the slow massive heavy quarks and the fast massless gluons suggest that
such systems may be amenable to a Born-Oppenheimer treatment, similar to
diatomic molecules\cite{hasenfratz}.  The slow heavy quarks correspond
to the nuclei in diatomic molecules, whereas the fast gluon and light-quark
fields correspond to the electrons.  One expects that the gluon/light-quark
wavefunctionals adapt nearly instantaneously to changes in the
heavy quark-antiquark separation.  At leading order, the gluons and
light quarks provide adiabatic potentials $V_{Q\bar{Q}}(r)$ which can be
computed in lattice simulations.  The leading order behavior of the heavy
quarks is then described by solving the Schr\"odinger equation separately
for each $V_{Q\bar{Q}}(r)$:
\begin{equation}
 \left\{ \frac{\bm{p}^2}{2\mu}+V_{Q\bar{Q}}(r)\right\}\Psi_{Q\bar{Q}}(r)
 = E\Psi_{Q\bar{Q}}(r),
\end{equation}
where $\mu$ is the reduced mass of the quark-antiquark pair and $r$
is the quark-antiquark separation.
The Born-Oppenheimer approximation provides a clear and unambiguous picture
of conventional and hybrid mesons: conventional mesons arise from the
lowest-lying adiabatic potential, whereas hybrid mesons arise from the
excited-state potentials.  The validity of such a Born-Oppenheimer picture
will be demonstrated in this talk.

\subsection{Excitations of the static quark potential}

The first step in a Born-Oppenheimer treatment of heavy quark mesons
is determining the gluonic terms $V_{Q\bar{Q}}(r)$.  Since familiar
Feynman diagram techniques fail and the Schwinger-Dyson equations
are intractable, the path integrals needed to determine $V_{Q\bar{Q}}(r)$
are usually estimated using Markov-chain Monte Carlo methods.
The spectrum of gluonic excitations in the presence of a static
quark-antiquark pair has been accurately determined in recent lattice 
simulations\cite{earlier1,earlier2,jkm2}
which make use of anisotropic lattices, improved actions, and large sets
of operators with correlation matrix techniques.

\begin{figure}[t]
\includegraphics[width=8cm,bb=50 50 554 770]{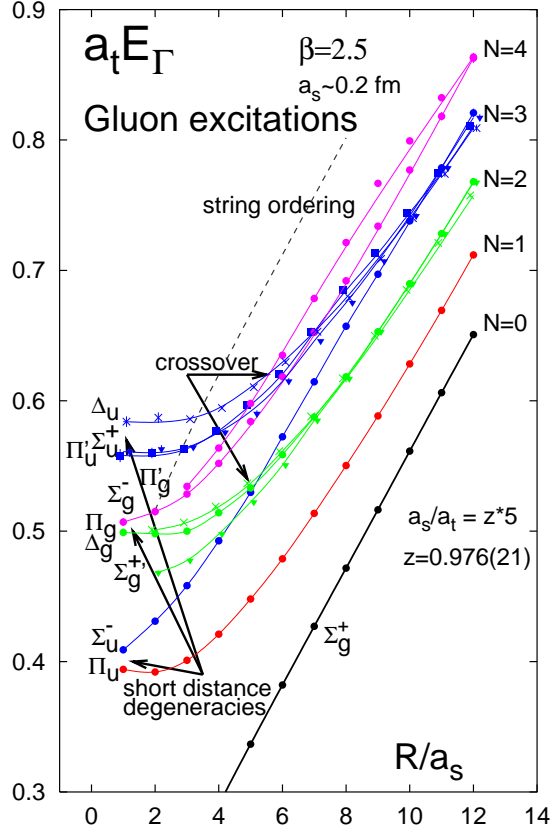}
\caption{ The spectrum of gluonic excitations in the presence of a static
 quark-antiquark pair from Ref.~\protect\cite{jkm2}.
 The solid curves are only shown for visualization.
 At short distances, the level orderings and degeneracies are
 consistent with the states expected in a multipole operator product
 expansion.  At large distances, the levels are consistent with
 the expectations from an effective string theory description.
 A dramatic level rearrangement is observed in the crossover
 region between $0.5-2.0$ fm. The dashed line marks a lower bound
 for the onset of mixing effects with glueball states which requires
 careful interpretation.
\label{fig:static_qqbar}}
\end{figure}

\begin{figure}[t]
\includegraphics[width=6cm,bb=0 0 730 518]{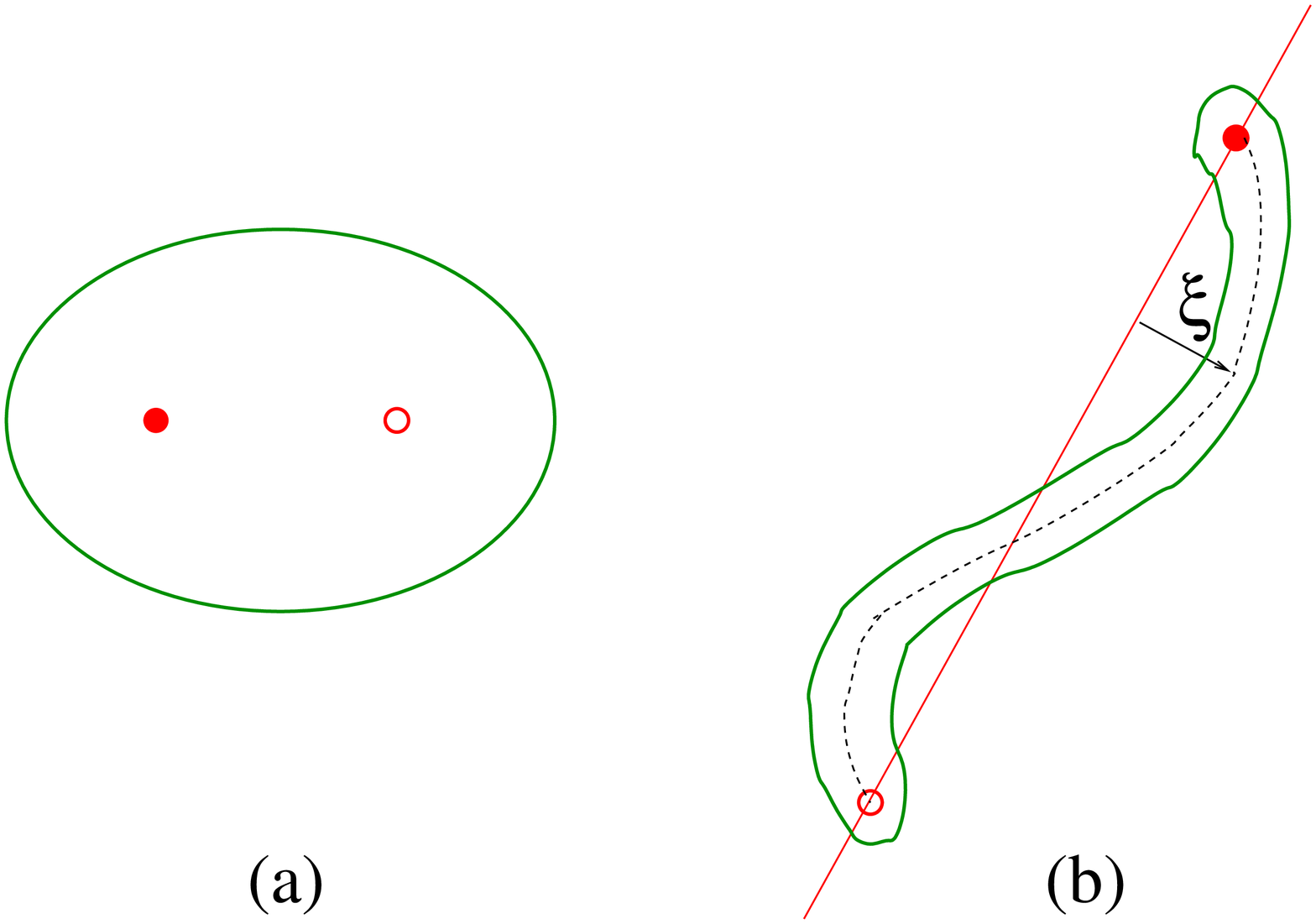}
\caption{One possible interpretation of the spectrum in 
 Fig.~\protect\ref{fig:static_qqbar}. (a) For small quark-antiquark
 separations, the strong chromoelectric field of the $Q\bar{Q}$ pair
 repels the physical vacuum (dual Meissner effect) creating a bubble.
 The low-lying stationary states are explained by the gluonic modes
 inside the bubble, since the bubble surface excitations are likely
 to be higher lying.
 (b) For large quark-antiquark separations, the bubble stretches into
 a thin tube of flux, and the low-lying states are explained by the
 collective motion of the tube since the internal gluonic excitations are
 much higher lying.
\label{fig:interp}}
\end{figure}

The results for one particular lattice spacing are shown in 
Fig.~\ref{fig:static_qqbar}.  Due to computational limitations, light quark
loops have been neglected in these calculations; their expected impact on
the meson spectrum will be discussed below.  The levels in 
Fig.~\ref{fig:static_qqbar} are labeled
by the magnitude $\Lambda$ of the projection of the total angular momentum
${\bf J}_g$ of the gluon field onto the molecular axis, and by $\eta=\pm 1$,
the symmetry under charge conjugation combined with spatial inversion
about the midpoint between the $Q$ and $\overline{Q}$.  States with
$\Lambda=0,1,2,\dots$ are denoted by $\Sigma, \Pi, \Delta, \dots$,
respectively.  States which are even (odd) under the above-mentioned
$CP$ operation are denoted by the subscripts $g$ ($u$).  An additional
$\pm$ superscript for the $\Sigma$ states refers to even or odd symmetry
under a reflection in a plane containing the molecular axis. 

These $V_{Q\bar{Q}}(r)$ potentials tell us much about the nature of
the confining gluon field between a quark and an antiquark.  Innumerable
lattice QCD simulations have confirmed the linearly rising ground-state
static quark-antiquark potential from gluon exchange.  Such a linearly
rising potential naively suggests that the gluon field forms a string-like
confining object connecting the quark and the antiquark.
However, it should be noted that the spherical bag model also predicts
a linearly rising potential for moderate $r$, 
and hence, the linearly rising ground-state
potential is {\em not} conclusive evidence of string formation.
Computations of the gluon action density surrounding a static quark-antiquark
pair in $SU(2)$ gauge theory also hint at flux tube formation\cite{bali_flux}.  

The spectrum shown in Fig.~\ref{fig:static_qqbar} provides
unequivocal evidence that the gluon field can be well approximated by an
effective string theory for large separations $r$. However,
string formation does not appear to set in until the quark and the antiquark
are separated by about 2 fm.  For small separations, the level orderings and
degeneracies are not consistent with the expectations from an effective string
description.  More importantly, the gaps differ appreciably from $N\pi/r$
with $N=1,2,3,\dots$.
Such deviations cannot be considered mere corrections, making the applicability
of an effective string description problematical.
Between 0.5 to 2 fm, a dramatic level rearrangement occurs.
For separations above 2 fm, the levels agree {\em without exception} with
the ordering and degeneracies expected from an effective string theory.
The gaps agree well with $N\pi/r$, but a fine structure remains.  The
$N\pi/r$ gaps are a robust prediction of any effective string theory since
they are a feature of the Goldstone modes associated with the spontaneous
breaking of transverse translational symmetry.  However, the details of the
underlying string theory are encoded in the fine structure.  This first
glimpse of such a fine structure offers the exciting possibility of
ultimately understanding the nature of the QCD string in future higher
precision simulations.

Fig.~\ref{fig:interp} illustrates one possible interpretation of the
results shown in Fig.~\ref{fig:static_qqbar}.  At small quark-antiquark
separations, the strong chromoelectric field of the $Q\bar{Q}$ pair
repels the physical vacuum in a dual Meissner effect, creating a bubble
surrounding the $Q\bar{Q}$.
The low-lying stationary states are explained by the gluonic modes
inside the bubble, since the bubble surface excitations are likely
to be higher lying.
 For large quark-antiquark separations, the
bubble stretches into a thin tube of flux, and the low-lying states
are explained by the collective motion of the tube since the internal
gluonic excitations, being typically of order 1 GeV, are now much higher
lying.

\subsection{The leading Born-Oppenheimer approximation}

In the leading Born-Oppenheimer approximation, one replaces
the covariant Laplacian $\bm{D}^2$ by an ordinary Laplacian $\bm{\nabla}^2$,
which neglects retardation effects.  The spin interactions of the 
heavy quarks are
also neglected, and one solves the radial Schr\"odinger equation:
\begin{equation}
 -\frac{1}{2\mu} \frac{d^2u(r)}{dr^2}
 + \left\{ \frac{\langle \bm{L}^2_{Q\bar{Q}}\rangle}{2\mu r^2}
   + V_{Q\bar{Q}}(r)\right\}u(r) = E\ u(r),
\end{equation}
where $u(r)$ is the radial wavefunction of the quark-antiquark pair.
The total angular momentum
is given by
\begin{equation}
 \bm{J}=\bm{L}+\bm{S},\quad \bm{S}=\bm{s}_Q + \bm{s}_{\bar{Q}},\quad
 \bm{L}=\bm{L}_{Q\bar{Q}}+\bm{J}_g,
\end{equation}
where $\bm{s}_Q$ is the spin of the heavy quark, $\bm{s}_{\bar{Q}}$ is the
spin of the heavy antiquark, $\bm{J}_g$ is the total spin of the gluon
field, and $\bm{L}_{Q\bar{Q}}$ is the orbital angular momentum of the
quark-antiquark pair.  In the LBO, both $L$ and $S$ are good quantum
numbers.  The expectation value in the centrifugal term is given by
\begin{equation}
  \langle \bm{L}^2_{Q\bar{Q}}\rangle=\langle \bm{L}^2\rangle
  - 2\langle \bm{L}\cdot\bm{J}_g\rangle+\langle\bm{J}_g^2\rangle.
\end{equation}
The first term yields $L(L+1)$.  The second term is evaluated by expressing
the vectors in terms of components in the body-fixed frame.  Let $L_r$
denote the component of $\bm{L}$ along the molecular axis, and $L_\xi$
and $L_\zeta$ be components perpendicular to the molecular axis.
Writing $L_\pm=L_\xi\pm iL_\zeta$ and similarly for $\bm{J}_g$, one obtains
\begin{equation}
\langle \bm{L}\cdot\bm{J}_g\rangle=\langle L_r J_{gr}\rangle
 +\textstyle\frac{1}{2}\langle L_+ J_{g-}+L_-J_{g+}\rangle.
\end{equation}
Since $J_{g\pm}$ raises or lowers the value of $\Lambda$, this term mixes
different gluonic stationary states, and thus, must be neglected in the
leading Born-Oppenheimer approximation.  In the meson rest frame,
the component of $\bm{L}_{Q\bar{Q}}$ along the molecular axis vanishes,
and hence, $\langle L_r J_{gr}\rangle=\langle J_{gr}^2\rangle=\Lambda^2$.
In summary, the expectation value in the centrifugal term is given in the 
adiabatic approximation by
\begin{equation}
  \langle \bm{L}^2_{Q\bar{Q}}\rangle=L(L+1)-2\Lambda^2
  +\langle\bm{J}_g^2\rangle.
\end{equation}
We assume $\langle\bm{J}_g^2\rangle$ is saturated by the minimum
number of allowed gluons.  Hence,
$\langle\bm{J}_g^2\rangle=0$ for the $\Sigma_g^+$ level and
$\langle\bm{J}_g^2\rangle=2$ for the $\Pi_u$ and $\Sigma_u^-$ levels.
Wigner rotations are used as usual to construct $\vert LSJM;\lambda\eta\rangle$
states, where $\lambda=\bm{J}_g\cdot\hat{\bm{r}}$ and $\Lambda=\vert\lambda\vert$,
then $J^{PC}$ eigenstates are finally obtained from
\begin{equation}
 \vert LSJM;\lambda\eta\rangle +\varepsilon\vert LSJM;-\lambda \eta\rangle,
\end{equation}
where $\varepsilon=1$ for $\Sigma^+$ levels, $\varepsilon=-1$ for $\Sigma^-$
levels, and $\varepsilon=\pm 1$ for $\Lambda\geq 1$ levels.
Hence, the $J^{PC}$ eigenstates satisfy
\begin{equation}
  P=\varepsilon (-1)^{L+\Lambda+1},\qquad C=\eta\varepsilon (-1)^{L+S+\Lambda}.
\end{equation}
Note that many levels are degenerate in the LBO approximation:
\[\begin{array}{rl}
 \Sigma_g^+(S): & 0^{-+}, 1^{--},\\[1mm]
 \Sigma_g^+(P): & 0^{++}, 1^{++}, 2^{++}, 1^{+-},\\[1mm]
 \Pi_u(P): & 0^{-+}, 0^{+-}, 1^{++}, 1^{--}, 1^{+-}, 1^{-+}, 2^{+-}, 2^{-+}.
\end{array}\]

\begin{figure}
\includegraphics[width=11cm, bb= 167 350 488 578]{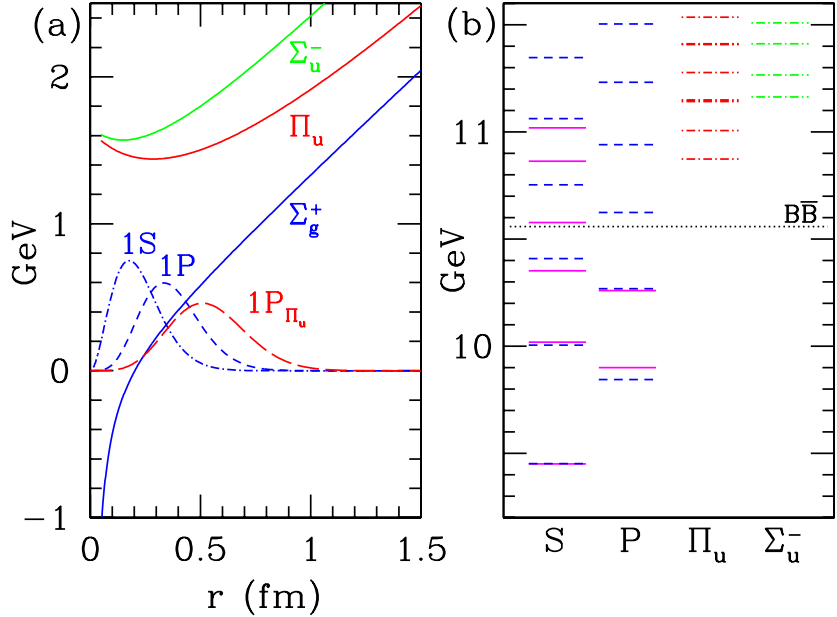}
\caption{(a) Static potentials and radial probability densities
 against quark-antiquark separation $r$
 for the conventional $1S$ and $1P$ bottomonium states and the hybrid
 $1P_{\Pi_u}$ level.
 The scale is set using $r_0^{-1}=450$ MeV.
 (b) Spin-averaged spectrum
 in the LBO approximation (light quarks neglected).  Solid lines
 indicate experimental measurements.  Short dashed lines indicate 
 the $S$ and $P$ state masses obtained using the $\Sigma_g^+$ potential
 with $M_b=4.58$ GeV. Dashed-dotted
 lines indicate the hybrid quarkonium states obtained from the $\Pi_u$
 $(L=1,2,3)$ and $\Sigma_u^-$ $(L=0,1,2)$ potentials.  These results
 are from Ref.~\protect\cite{jkm}.
\label{fig:LBO}}
\end{figure}

\begin{figure}
\includegraphics[width=11cm, bb=0 340 596 730]{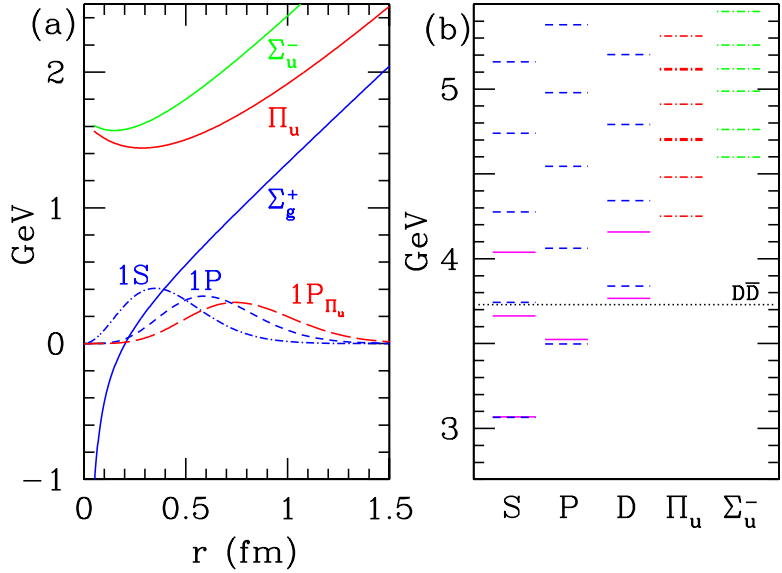}
\caption{(a) Static potentials and radial probability densities
 against quark-antiquark separation $r$ with $r_0^{-1}=415$ MeV
 for the conventional $1S$ and $1P$ charmonium levels and the
 hybrid $1P_{\Pi_u}$ state.
 (b) Spin-averaged spectrum
 in the LBO approximation (light quarks neglected).  Solid lines
 indicate experimental measurements.  Short dashed lines indicate 
 the $S, P,$ and $D$ state masses obtained using the $\Sigma_g^+$ potential
 with $M_c=1.20$ GeV. Dashed-dotted
 lines indicate the hybrid quarkonium states obtained from the $\Pi_u$
 $(L=1,2,3)$ and $\Sigma_u^-$ $(L=0,1,2)$ potentials.  
\label{fig:LBOcc}}
\end{figure}

The LBO spectrum\cite{jkm} of conventional $\overline{b}b$
and hybrid $\overline{b}gb$ states is shown in Fig.~\ref{fig:LBO}. 
Below the $\overline{B}B$ threshold, the LBO results agree
well with the spin-averaged experimental measurements of bottomonium
states (any small discrepancies disappear once light quark loops are
included).   Above the threshold, agreement with experiment is lost, suggesting
significant corrections either from mixing and other higher-order effects or
(more likely) from light sea quark effects.  Note from the radial probability
densities shown in Fig.~\ref{fig:LBO} that the size of the hybrid state is
large in comparison with the conventional $1S$ and $1P$ states. 
The analogous results in charmonium are shown in Fig.~\ref{fig:LBOcc}.

\begin{figure}
\includegraphics[width=10cm,bb= 28 144 572 552]{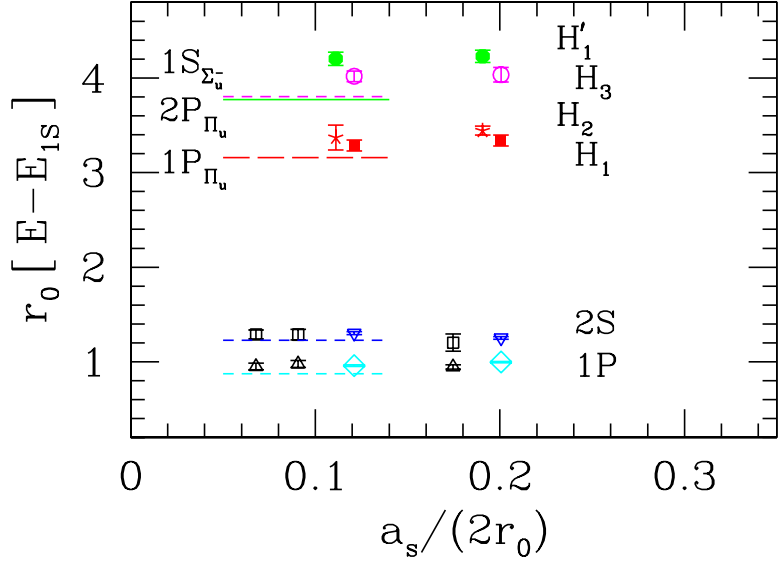}
\caption{
  Simulation results from Ref.~\protect\cite{jkm} for the heavy quarkonium
  level splittings of two conventional levels and four hybrid levels
  (in terms of $r_0$ and with respect to the $1S$ state) against the lattice
  spacing $a_s$. Results from Ref.~\protect\cite{Wilson} using an NRQCD
  action with higher-order corrections are shown as open boxes and
  $\bigtriangleup$.  The horizontal lines show the LBO predictions.
  Agreement of these splittings within 10\% validates the leading 
  Born-Oppenheimer approximation (in the absence of light quarks).
\label{fig:scaling}}
\end{figure}

The validity of such a simple physical picture relies on the smallness
of higher-order spin, relativistic, and retardation effects, as well as
mixings between states based on different $V_{Q\bar{Q}}(r)$.  
The importance of retardation and leading-order mixings between states
based on different adiabatic potentials was tested in Ref.~\cite{jkm}
by comparing the LBO level splittings with those determined from meson
simulations using a leading-order non-relativistic (NRQCD) heavy-quark
action.  The NRQCD action included only a covariant temporal derivative
and the leading covariant kinetic energy operator; quark spin and
$\bm{D}^4$ terms were neglected.  The simulations include retardation
effects and allow possible mixings between different adiabatic surfaces.
The level splittings (in terms of the hadronic scale $r_0$ and with respect
to the $1S$ state) of the conventional $2S$ and $1P$
states and four hybrid states were compared (see Fig.~\ref{fig:scaling})
and found to agree within $10\%
$, strongly supporting the validity of the
leading Born-Oppenheimer picture.   The operators used 
to create the mesons in the simulations
are described in Table~\ref{table:mesonops}.

\begin{table}[t]
\renewcommand{\arraystretch}{1.2}  
\renewcommand{\tabcolsep}{8mm}
\begin{tabular}{clcc}\hline
 $J^{PC}$ & &  
 \tablehead{1}{c}{b}{Degeneracies} & 
 \tablehead{1}{c}{b}{Operator} \\ \hline 
 $0^{-+}$ & $S$ wave & $1^{--}$ &
    $\chi^\dag\ (\bm{D}^2)^p\ \psi$ \\
 $1^{+-}$ & $P$ wave & $0^{++},1^{++},2^{++}$ &$\chi^\dag
    \ \bm{D} \ \psi$ \\
 $1^{--}$ & $H_1$ hybrid & $0^{-+},1^{-+},2^{-+}$ &
    $\chi^\dag\ \bm{B} (\bm{D}^2)^p\ \psi$ \\
 $1^{++}$ & $H_2$ hybrid & $0^{+-},1^{+-},2^{+-}$ & $\chi^\dag\ 
      \bm{B}\!\times\!\bm{D}
      \ \psi$ \\
 $0^{++}$ & $H_3$ hybrid & $1^{+-}$ & $\chi^\dag\ \bm{B}
     \!\cdot\!\bm{D}\ \psi$ \\ \hline
\end{tabular}
\caption{
  The meson spin-singlet operators used
  in the simulations of Ref.~\protect\cite{jkm} in terms of the
  heavy quark two-component field $\psi$, antiquark field $\chi$,
  covariant derivative $\bm{D}$, and chromomagnetic field $\bm{B}$.
  Note that $p=0,1,2,$ and $3$
  were used to produce four distinct operators in the $0^{-+}$
  and $1^{--}$ sectors.  In the third column are listed the spin-triplet
  states which can be formed from the operators in the last column; the
  states in each row are degenerate for the NRQCD action used here.}
 \label{table:mesonops}
\end{table}

\subsection{Quark spin effects and light-quark loops}

\begin{figure}
\includegraphics[width=7cm,bb= 49 162 515 611]{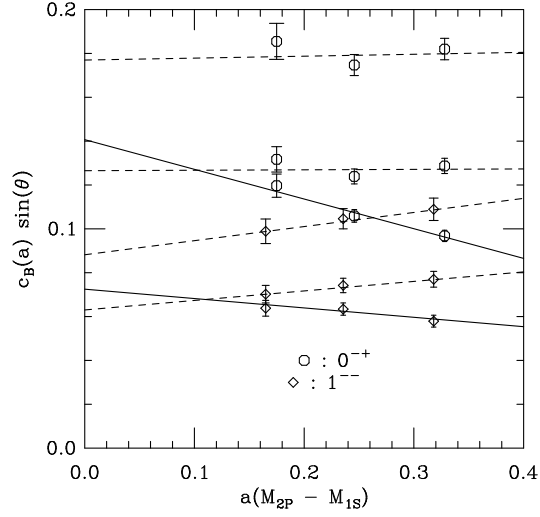}
\caption{ Hybrid/$S$-wave configuration mixing angle against lattice
 spacing. For each channel, the lowest three points with solid fit line
 reflect tree-level values for $c_B$, and the higher two sets of
 three points result from setting $c_B$ using 30 and 60 MeV
 for the $S$-wave hyperfine splitting. 
\label{fig:toussaint}}
\end{figure}

A very recent study\cite{burch} has shown that heavy-quark spin effects
are unlikely to spoil the Born-Oppenheimer approximation.  Using
lowest-order lattice NRQCD to create heavy-quark propagators, a basis
of unperturbed $S$-wave and $\vert 1H\rangle$ hybrid states were formed.  
The $c_B \bm{\sigma\!\cdot\! B}/2M_Q$ spin interaction was then applied
at an intermediate time slice to compute the mixings between such
states due to this interaction in the quenched approximation
(see Fig.~\ref{fig:toussaint}).
For a reasonable range of $c_B$ values, the following results were
obtained:
\[\begin{array}{lcl} 
 \langle 1H\vert\Upsilon\rangle &\approx& 0.076-0.11, \\
 \langle 1H\vert\eta_b\rangle &\approx& 0.13-0.19, \end{array}
\qquad
\begin{array}{lcl} 
 \langle 1H\vert J/\Psi\rangle &\approx& 0.18-0.25, \\
 \langle 1H\vert\eta_c\rangle &\approx& 0.29-0.4. \end{array}
\]
Hence, mixings due to quark spin effects in bottomonium are very
small, and even in charmonium, the mixings are not large.

\begin{figure}
\includegraphics[width=7cm,bb= 50 50 554 554 ]{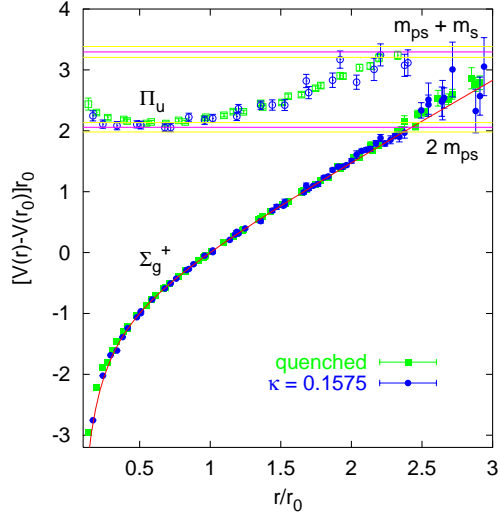}
\caption{ Ground $\Sigma_g^+$ and first-excited $\Pi_u$
 static quark potentials without sea quarks (squares, quenched) and
 with two flavors of sea quarks, slightly lighter than the strange
 quark (circles, $\kappa=0.1575$).  Results are given in terms of
 the scale $r_0\approx 0.5$ fm, and the lattice spacing is 
 $a\approx 0.08$ fm. Note that $m_S$ and $m_{PS}$ are the masses
 of a scalar and pseudoscalar meson, respectively,
 consisting of a light quark and a static antiquark.
 These results are from Ref.~\protect\cite{SESAM}.
\label{fig:hybridpot}}
\end{figure}

In the absence of light quark loops, one obtains a very dense spectrum of
mesonic states since the $V_{Q\bar{Q}}(r)$ potentials increase indefinitely
with $r$.  However, the inclusion of light quark loops changes the
$V_{Q\bar{Q}}(r)$ potentials.  First, there are slight corrections at small
$r$, and these corrections remove the small discrepancies of the LBO
predictions with experiment below the $B\overline{B}$ threshold seen in
Fig.~\ref{fig:LBO}.  For large $r$, the inclusion of light quark loops
drastically changes the behavior of the $V_{Q\bar{Q}}(r)$ potentials:
instead of increasing indefinitely, these potentials eventually
level off at a separation above 1 fm when the static quark-antiquark
pair, joined by gluonic flux, can undergo fission into 
$(Q\overline{q})(\overline{Q}q)$, where $q$ is a light quark.  Clearly,
such potentials cannot support the populous set of states shown in
Fig.~\ref{fig:LBO}; the formation of bound states and resonances
substantially extending over 1~fm in diameter seems unlikely. 
A complete open-channel calculation taking the effects of including the
light quarks correctly into account has not yet been done, but unquenched
lattice simulations\cite{SESAM} show that the $\Sigma_g^+$ and $\Pi_u$
potentials change very little for separations below 1 fm when sea
quarks are included (see Fig.~\ref{fig:hybridpot}), suggesting that a 
handful of low-lying states whose wavefunctions do not extend appreciably
beyond 1 fm in diameter may exist as well-defined resonances in nature.

\subsection{Simulations with relativistic heavy quarks}

\begin{figure}
\includegraphics[width=7cm,bb= 26 56 507 434]{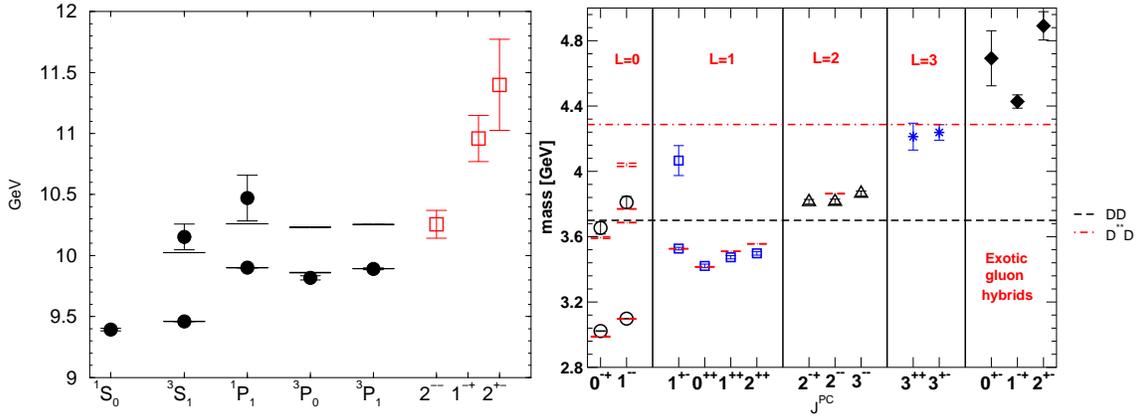}
\includegraphics[width=8cm,bb=46 37 773 561]{cchybrids}
\caption{ The bottomonium spectrum (left) from Ref.~\protect\cite{manke1}
 and charmonium spectrum (right) from Ref.~\protect\cite{manke2}
 in the quenched approximation using an anisotropic clover fermion
 action.  The $^1P_1-^3S_1$ splitting is used to set the scale.  Experimental
 values are indicated by the horizontal lines.  In the $b\overline{b}$ spectrum,
 the two rightmost points indicate exotic hybrid candidates which agree with the
 Born-Oppenheimer predictions, but with very large uncertainties. In the
 $c\overline{c}$ spectrum, the three rightmost
 points indicate results for hybrid candidates.
\label{fig:bbhybrids}}
\end{figure}

A recent quenched calculation\cite{manke1} of bottomonium hybrids using a
relativistic heavy-quark action on anisotropic lattices confirms the
predictions of the Born-Oppenheimer approximation, but admittedly,
the uncertainties in the simulation results are large 
(see Fig.~\ref{fig:bbhybrids}).  These calculations make use of a
Symanzik-improved anisotropic gauge action and an improved fermion
clover action.  These same authors have also recently studied the
charmonium spectrum\cite{manke2}.  The results,
shown in Fig.~\ref{fig:bbhybrids},
suggest significant, but not large, corrections to the
leading Born-Oppenheimer approximation.

\section{Other tidbits}

Some other recent studies involving gluonic excitations are summarized
in the remainder of this talk.

\subsection{Glueballs}

\begin{figure}[t]
\includegraphics[width=7cm,bb= 11 98 535 631]{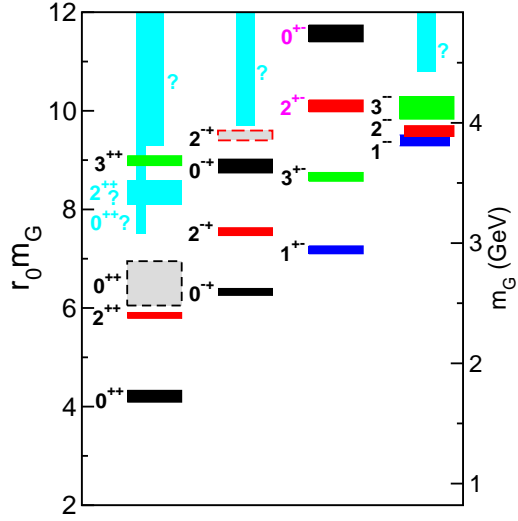}
\caption{ The mass spectrum of glueballs in the pure SU(3) gauge theory
 from Ref.~\protect\cite{glueballs}.
 The masses are given in terms of the hadronic scale $r_0$ along the
 left vertical axis and in terms of GeV along the right vertical axis
 (assuming $r_0^{-1}=410(20)$ MeV).  The mass uncertainties indicated by the
 vertical extents of the boxes do {\em not} include the uncertainty in
 setting $r_0$.  The locations of states whose interpretation requires
 further study are indicated by the dashed hollow boxes.  The shaded
 strips with accompanying question marks indicate regions in which the
 spectrum is not known.
\label{fig:contglue}}
\end{figure}

\begin{figure}
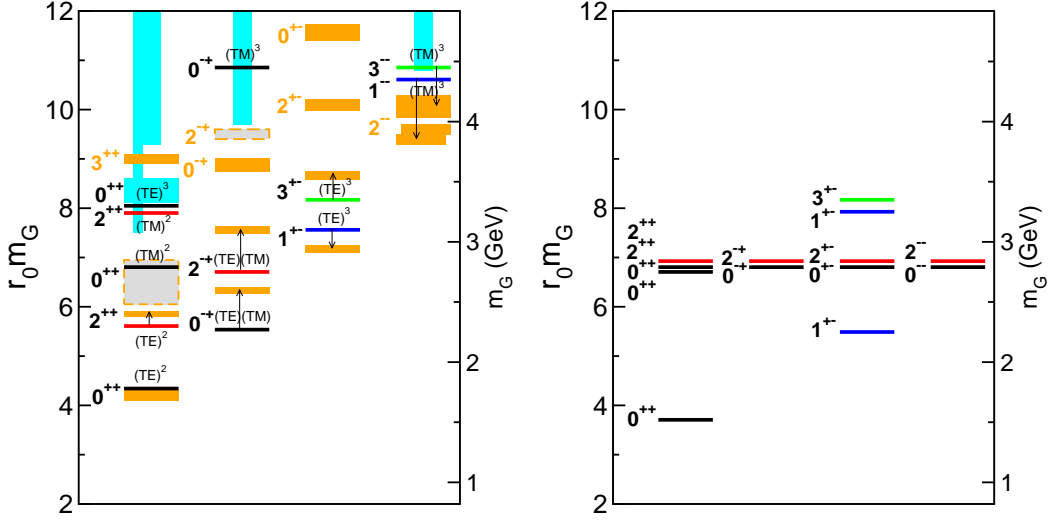

\includegraphics[width=7cm,bb=11 98 535 631]{bag_glueballs}
\includegraphics[width=7cm,bb=11 98 535 631]{ftube_glueballs}
\caption{ Comparison of the glueball spectrum obtained from Monte Carlo
 simulations with that predicted by the bag model (left) with
 $\alpha_s=0.5$ and $B^{1/4}=280$ MeV, and the
 Isgur-Paton flux tube model (right).  The crude bag model appears to
 capture the qualitative features of the spectrum, whereas the flux
 tube model fails entirely.  The bag model results shown above do not
 include gluon self-energies, but the incorporation of such contributions
 is not expected to change the spectrum appreciably.
\label{fig:glueballs_models}}
\end{figure}

The glueball spectrum in the absence of virtual quark-antiquark
pairs is now well known\cite{glueballs} and is shown in
Fig.~\ref{fig:contglue}.  The glueball spectrum
can be qualitatively understood in terms of the interpolating
operators of minimal dimension which can create glueball 
states\cite{jaffe} and can be reasonably well explained\cite{kuti}
in terms of a simple constituent gluon (bag) model which approximates
the gluon field using spherical cavity Hartree modes with residual 
perturbative interactions\cite{gluebag1,gluebag2}, as shown in
Fig.~\ref{fig:glueballs_models}.  This figure also shows that a 
model\cite{paton} of glueballs as loops of chromoelectric flux
completely fails to explain the spectrum.

\subsection{Light-quark exotic $1^{-+}$ hybrid meson}

\begin{figure}
\includegraphics[width=9cm,bb=58 127 580 655]{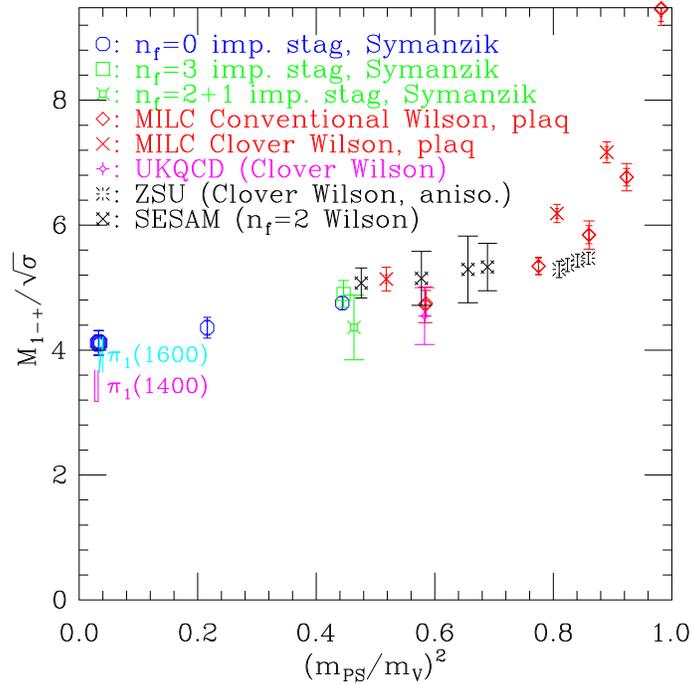}
\caption{Summary of recent determinations of the $1^{-+}$ hybrid meson
 mass against $(m_{\rm PS}/m_{\rm V})^2$.  The bold octagon indicates
 a linear extrapolation of quenched $n_f=0$ results to the physical
 point $(m_{\rm PS}/m_{\rm V})^2=0.033$.  The most recent results make
 use of improved staggered fermions\protect\cite{KShybrid} and remain
 somewhat heavier than the experimental candidates.
 } 
\label{fig:lighthybrids}
\end{figure}

There are recent new determinations of the exotic $1^{-+}$
meson mass\cite{KShybrid} using improved staggered fermions and
the Wilson gluon action at a lattice spacing $a\approx 0.09$ fm.
Both quenched $(n_f=0)$ and unquenched $(n_f=2+1,3)$ simulation results
are presented.  The $n_f=3$ and $n_f=2+1$ simulation differ in how the
three quark mass parameters $m_u$, $m_d$, and $m_s$ were chosen.
In the $n_f=3$ simulations, $m_u=m_d=m_s$ near the
strange quark mass was used, and in the $n_f=2+1$ simulations,
$m_u=m_d=0.4m_s$ was used.  The results are compared to previous
determinations in Fig.~\ref{fig:lighthybrids} and remain somewhat heavier
than the experimental candidates.

\subsection{Static three-quark potential}

\begin{figure}
\includegraphics[width=9cm,bb=0 0 288 288]{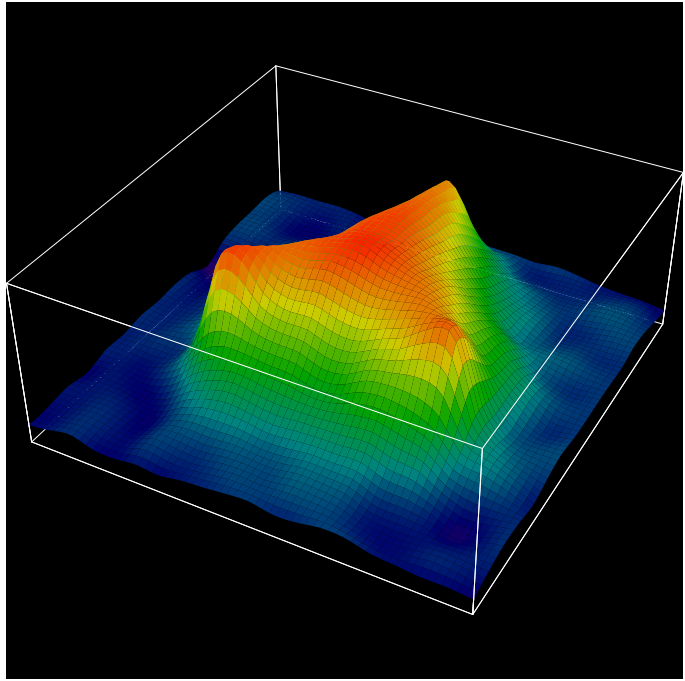}
\caption{The abelian action density of gluons in the presence of three
 static quarks from Ref.~\protect\cite{ichie}.  These results support
 the so-called $Y$ ansatz, and were obtained using the Wilson gauge action
 with $\beta=6.0$ (lattice spacing $a\sim 0.1$ fm)
 on a $16^3\times 32$ lattice.  The three static quarks
 were located at sites $(17,14), (22,6)$, and $(12,6)$ in the $xy$-plane.
 A calculation in Ref.~\protect\cite{ichie2} including light quark
 loops shows similar results.
 } 
\label{fig:threequark}
\end{figure}

The behavior of gluons in the presence of three static quarks has come
under recent study and promises to shed light on the structure of
baryons.  In Ref.~\cite{ichie}, the abelian action density of gluons
in the presence of three static quarks has been calculated and is
shown in Fig.~\ref{fig:threequark}.  The results favor a $Y$-ansatz in
which the quarks are confined by a genuine three-body force consisting
of three gluonic fluxes meeting at a common junction.  An alternative
$\Delta$-ansatz composed of three sets of two-body interactions
is disfavored.  A calculation in Ref.~\cite{ichie2} which includes light
sea quarks shows similar results.

\begin{figure}
\includegraphics[width=9cm,bb=32 18 530 440]{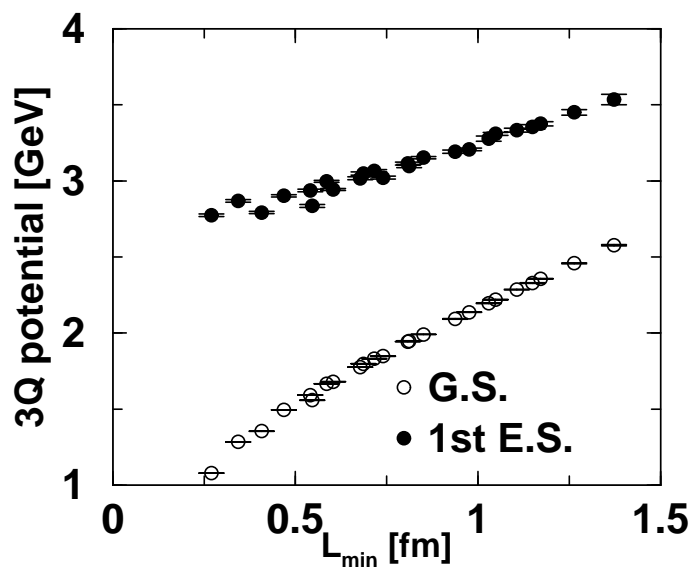}
\caption{Energies of the ground and first-excited stationary states
 of gluons in the presence of three static quarks against
 $L_{\rm min}$, the minimal total length of three line segments connecting
 the three quarks at a common junction. The scale is
 set using the string tension $\sigma=0.89$ GeV/fm.  
 These results are from Ref.~\protect\cite{suganuma} and
 were obtained using the Wilson gauge action
 at $\beta=5.8$ on a $16^3\times 32$ lattice.} 
\label{fig:threequarkB}
\end{figure}

The first excitation of gluons in the presence of three static quarks
has also recently been studied\cite{suganuma}.  The results are shown
in Fig.~\ref{fig:threequarkB}.  Although systematic uncertainties
from finite lattice spacing and finite volume still need to be
investigated, the results indicate an excitation energy near 1 GeV.

\section{Conclusion}
Hadronic states bound by an excited gluon field are an interesting
new form of matter.  Theoretical investigations into their nature must
confront the long-standing problem of understanding the confining
gluon field, but for this reason, such states are a potentially rich
source of information concerning quark confinement in QCD.

The study of heavy-quark mesons is a natural starting point in the
search to understand hadron formation.  The Born-Oppenheimer approximation
provides a compelling and clear physical picture of both conventional and
hybrid heavy-quark mesons.  In this talk, the evidence
supporting the validity of a Born-Oppenheimer treatment of
such systems was presented.  The validity of the leading Born-Oppenheimer
approximation in the absence of light quarks was established by a comparison
of LBO level splittings with direct simulation results.  Mixings induced
by quark spin effects were then shown to be small, and the effects of
including light sea quarks were discussed, suggesting that a handful
of hybrid mesons with diameters not extending appreciably beyond 1 fm
may exist as well-defined resonances in nature.

Other recent studies involving gluonic excitations were also summarized.
In particular, the pure-gauge glueball masses, the light quark
$1^{-+}$ hybrid meson mass, and the static three-quark potential and its
first excitation were presented. This work was supported by the 
U.S.\ National Science Foundation under award PHY-0099450,
the U.S.\ DOE, Grant No. DE-FG03-97ER40546, and the European
Community's Human Potential Programme under contract HPRN-CT-2000-00145,
Hadrons/Lattice QCD.

\bibliographystyle{aipprocl}
\bibliography{HQhybrids}
\end{document}